# The hidden symmetry and Mr. Higgs!

*Costas J. Papachristou* *

***Abstract.*** Written in non-technical language, this review article explains the significance of the Higgs field and the associated Higgs boson in High-Energy Physics. The connection of symmetry with particle interactions and their unification is also discussed in this context. The presentation is informal and physical concepts are demonstrated through metaphors from everyday experience.

## *1. Introduction*

One of the dominant scientific issues in 2012 and 2013 was the experiments that took place at the **CERN** research center in Geneva. Their major goal was the experimental verification of the existence of a mysterious particle which constitutes a fundamental ingredient of the model that we believe describes the elementary building blocks of matter and the interactions among them. The **Higgs boson** (the quantum of the **Higgs field**) was indeed the biggest bet of the research efforts, and the verification of its existence – pending some remaining issues, to be resolved in the near future – was hailed as a triumph of High-Energy Physics and, predictably, led to long-awaited and well-deserved Nobel prizes [1].

But, why was this elusive particle so important as to justify spending several billion dollars for its hunt at a time of economic world-crisis? Well, if not for anything else, perhaps for a deep sigh of relief of physicists – at least those who didn't bet on the collapse of modern High-Energy Physics in order to be given the historic chance of building it from the start!

This article constitutes an attempt to explain, in the simplest terms possible, the reason why the Higgs field and the associated Higgs particle are such important elements of contemporary physics theories that try to unlock the mysteries of the world that surrounds us. And, given that the matter we observe, at the most fundamental level, is made of **elementary particles** [2] (such as, e.g., the familiar *electron*, as well as others "residing" in the atomic nucleus), we begin our story by examining the different ways these particles interact with one another...

## *2. The hidden simplicity of Nature*

In accordance with the phenomenology of the low-energy world we live in, we can distinguish four kinds of **forces** (or **interactions**) [2] among the elementary constituents of matter:

(1) **Gravitational forces**, which are responsible for the weight of all objects, as well as for the motion of the Earth around the Sun.

* Department of Physical Sciences, Naval Academy of Greece
 papachristou@snd.edu.gr

(2) **Electromagnetic forces** (a force of such origin is, e.g., the friction we feel when we rub our hands against one another).

(3) **Strong forces**, to which the atomic nucleus owes its coherence despite the repulsion between the positively charged protons.

(4) **Weak forces**, responsible for a number of processes taking place inside the nucleus.

There are indications, however, that Nature is much simpler than it appears to be! For example, prior to the systematic theoretical formulation of the laws of electromagnetism by **James Clerk Maxwell** (1831-1879), electricity and magnetism were treated as two distinct physical phenomena, independent of each other. This was due, in part, to the apparent differences in the properties of electric and magnetic forces.

With his complex mathematical equations [3,4], Maxwell described the electric and the magnetic field as "two sides of the same coin", since each field may transform into the other, depending on the way we observe it. Hence, instead of two separate fields (electric and magnetic), we now speak of a single **electromagnetic field**.

It is interesting to note that (see, e.g., p. 588 of [5]), with regard to their relative strengths, the electric and the magnetic force become equivalent to each other in the limit of high speeds – thus, high energies – of the interacting electrical charges. This is a first hint that *the simplicity of Nature reveals itself to us only when sufficient energy is spent for its experimental observation!*

One of the biggest achievements of twentieth-century Physics was the discovery that, in a similar way, the electromagnetic and weak forces also represent two manifestations of a single interaction, the **electroweak force**. The problem of an even larger unification incorporating the strong interaction as well remains an open challenge. Gravity, on the other hand, is a different kind of problem since, in contrast to the other forces, it doesn't lend itself easily to a quantum formulation (see, e.g., [6]).

From the experimental point of view, the thing to keep in mind is that, as mentioned above, the simpler Nature appears to be through these successive stages of unification, the more expensive is the "ticket" the spectator of this simplicity is required to pay. And the name of this ticket is *energy*! That is, the supposed simplicity of Nature can only be revealed through very-high-energy experiments. And, the greater is the degree of simplicity, the more is the energy required. This explains the enormous expenditure for the construction of bigger and bigger elementary-particle accelerators, like the **Large Hadron Collider** (LHC) at CERN [7].

## 3. *The other side of the hill*

A simple example may help us better understand the situation: Imagine you reside at the foot of a hill located at the center of a town, the houses of which are exactly similar to each other and are uniformly distributed around the hill. From the point you are located, only a part of the town is visible since the hill blocks the view to the other side. Thus, from your point of view, there is "your" neighborhood and some other one, at the opposite side of the hill. At the point where you stand, your perception of the town is partial and *asymmetric*.

Suppose now you find the strength (that is, the required energy) to walk up to the top of the hill. From there you can look around and see every neighborhood of the town. The view is now complete and perfectly *symmetric* (no matter how you turn your body, you will always see some part of the town and, according to our assumption, all parts look alike). What we must keep in mind is that, *moving from the complexity of asymmetry toward the simplicity of symmetry requires the expenditure of energy!*

## 4. *The symmetry behind the interaction*

The elementary particles and the interactions (forces) among them are described by the so-called **Standard Model** [2,8]. This model is basically a synthesis of all experimentally verified theories on the structure of matter at the most fundamental level. An issue in need of experimental verification was the mechanism by which the particles (and, macroscopically, matter itself) acquire **mass** (or, if you prefer, **inertia**). Well, you may ask, isn't mass an inherent property of each particle, endowed to the particle from the very beginning of its creation? To understand the problem, it is necessary to go back to the concept of symmetry...

In the microworld, symmetry is much more than just a matter of aesthetics! Among other things, it is the factor that determines the kind of interaction between particles. That is, *behind every form of interaction there is a corresponding symmetry*, where by "symmetry" we mean **invariance** of some sort under certain mathematical transformations (see Appendix). As an example, the electromagnetic interaction between electrically charged particles can be associated with the symmetry (invariance in form) of the fundamental equations of Electromagnetism under specific abstract mathematical transformations of the functions that describe the electromagnetic field and the particles interacting through it [2].

According to quantum theory, the electromagnetic field itself is represented by its own "particles", **photons**. We can think of them as little spheres of energy exchanged between charged particles, making one particle aware of the presence of another. Photons are the **quanta** (the most elementary quantities) of the electromagnetic field. Their role is to communicate the electromagnetic interaction between electrically charged particles.

With regard to symmetry, the photon plays the role of a "messenger" who informs every observer by whom it passes about the details of the mathematical transformations performed on the functions representing the particles at neighboring points of space (or, more correctly, of *spacetime*).

A serious constraint, however, must be taken into account: The theories associating particle interactions with underlying symmetries demand that the quanta of the field responsible for an interaction should have **zero mass** [2]. This is indeed true for the photons (carriers of the electromagnetic interaction) but not for the quanta of the field associated with the weak interaction. Thus, the latter interaction would be at risk of staying out of the game of symmetry, and the theory of the unification of the weak force with the electromagnetic (electroweak force) would break down... if a mysterious field weren't there to save the game!

### 5. *The boring professor and his popular escort !*

The mass problem is dealt with by introducing the **Higgs field**. This field allows us to regard the quanta of all interactions as *intrinsically* massless. Their *apparent* property known as mass is due to their interaction with the Higgs field, or, if you prefer, with the quantum of this field, the **Higgs boson** [2,9]. (The term *"boson"* refers to particles with the property that any number of them can occupy the same quantum state. This is a fundamental property of the quanta of all interactions. This is *not* the case, however, with electrons or other *matter* particles such as *neutrinos* or *quarks* [2]!) Generally speaking, as proposed by **Peter Higgs** and other theoretical physicists working independently, the mass of any elementary particle is an *acquired* property that originates from the particle's interaction with the ubiquitous Higgs field.

Thus, hypothetically, if someone suddenly "turned off" this field (as we assume the case was for a small period after the Big Bang [6,10] due to extreme temperatures), all particles would appear *massless* (they would have no inertia, that is, they would not resist any attempt to alter their state of motion). According to the Theory of Relativity, this would mean that every particle would travel at the speed of light. We know, of course, that this isn't true in reality (with the exception of the photon).

Again, an example will be helpful: Imagine a ball organized by university students. In the big dancing room, a large number of students are uniformly distributed all over the place. Let us suppose that this multitude of students constitutes the "Higgs field" and each individual student represents a "Higgs boson" (a quantum of the field).

At some point in the evening, a boring professor (say, the author) makes his appearance at the ball. As he gets little attention upon entering the room, he can move more or less freely and accelerate almost at will. He is a "particle" with a small mass (a small inertia) since the Higgs field and its quanta (the students) do not bother much to slow down his motion!

Imagine now the late arrival of the professor's beautiful lady escort. As she attracts the attention of the students, they all rush to approach her, thus making it difficult for her to move inside the room. So, in order to speed up her step she will need to exert force: the Higgs field (the students) endowed her with a large mass (inertia). (Since this is only an allegoric paradigm, it should not be concluded that the lady is, literally, overweight!)

Now, if the students somehow became invisible, then an external observer might *assume* that this inertia is an inherent property of the woman. In a similar spirit, we presume that the inertia exhibited by all bodies is not an intrinsic property but simply a result of their interactions with the "invisible" (under normal, low-energy conditions) Higgs field. And this field becomes "visible" through its quantum, the Higgs boson.

## *6. Epilogue*

So, the latest experiments appear to confirm the Higgs theory, although several issues remain open and are in need of further investigation [11]. The delay in the discovery of the Higgs boson was due to the fact that this particle is extremely heavy, as it interacts strongly with its own field! Thus, its creation in the laboratory demands very high energies (remember the famous Einstein relation according to which mass and energy are equivalent). It was to this end that the LHC was built at CERN.

Physicists can now rejoice at the happy outcome of this enormous scientific endeavor, which – on top of everything – was excessively costly at a time of international economic crisis. What would be the consequences had these experiments failed to verify one of the fundamental predictions of the Standard Model? Well, a large part of Particle Physics as we know it would probably have to be revised and new approaches would have to be considered. It must be said, however, that, for some physicists such a scenario wouldn't necessarily be catastrophic! Any scientific theory is good for as long as it is supported by experiment. The experimental overturn of a theory, unpleasant as it may be, opens new paths and creates new opportunities in scientific research. Isn't this what happened, in an almost cataclysmic way, at the beginning of last century?

## *Acknowledgment*

I thank my students for penetrating questions that prompted me to write this article.

## *Appendix*

In Physics, a *field* is the assignment of a definite value to a physical quantity, for each point of spacetime. Thus, for example, the electromagnetic field is represented by a pair of vectors (***E***, ***B***), each of which takes on a certain value at each spacetime point ($x,y,z,t$). The manner in which these vectors change in space and time is described by a set of *differential equations*, called *Maxwell's Equations* [3-5]. In general, every field is associated with a corresponding differential equation (or set of differential equations) such that the field (viewed as a mathematical function) is a *solution* to this equation.

A transformation of a field leaving the corresponding differential equation *invariant in form* is said to represent a *symmetry* of this equation. Thus, a symmetry transformation produces a new solution of the field equation from any given solution. The fields that represent particle interactions emerge by demanding that the field equations for the interacting particles be *invariant* under certain groups of local transformations. (To be accurate, this invariance concerns the *Lagrangian function* associated with these equations.)

For more details on symmetries of differential equations, in general, the reader is referred to [12] and the extensive references therein.